\documentstyle[preprint,epsfig,aps]{revtex}
\def\al{\alpha} 
\def\be{\beta} 
\def\de{\delta} 
\def\la{\lambda} 
\def\nl{\nonumber\\}
\def\bea{\begin{eqnarray}} 
\def\eea{\end{eqnarray}} 
\newcommand{\wt}{\widetilde}
\input epsf.tex \def\DESepsf(#1 width #2){\epsfxsize=#2 \epsfbox{#1}}
\begin{document}
\preprint{\vbox{\hbox{}}} 
\draft 
\title{ Two-loop
Barr-Zee type Contributions \\  to $(g-2)_\mu$ in the MSSM
}
\author{Abdesslam Arhrib\footnote{On leave of absence from Department of 
Mathematics, FSTT, B.P 416, Tangier Morocco, arhrib@phys.ntu.edu.tw} 
and Seungwon Baek\footnote{swbaek@phys.ntu.edu.tw}} 
\address{ Department of
Physics, National Taiwan University,  Taipei, Taiwan }

\tighten

\date{\today} \maketitle
\begin{abstract}
We consider the contribution of a two-loop Barr-Zee type diagram to
$(g-2)_\mu$ in the minimal supersymmetric standard model (MSSM). 
At relatively  large  $\tan\beta$, we show that  
the contribution of light third generation scalar fermions and
neutral CP-even Higgs, $h^0(H^0)$, can easily explain
the very recent BNL experimental data.  
In our analysis $(g-2)_\mu$ prefers negative $A_{f}$ and positive $\mu$. 
It is more sensitive to the chirality flipping $h^0(H^0)\wt{f}_R^*\wt{f}_L$
rather than chirality conserving couplings.
\end{abstract}

\pacs{}

\preprint{\vbox{\hbox{}}}

\section{Introduction}

The recent measurement of the anomalous magnetic dipole moment (MDM) of
a muon, $a_\mu =(g_\mu-2)/2$, by Brookhaven E821 collaboration,
showed a $2.6\sigma$ deviation from the standard model (SM) calculation.
\bea
  \de a_\mu \equiv a_\mu^{\rm exp} -a_\mu^{\rm SM} 
     = (43 \pm 16) \times 10^{-10}.
\eea
Although the experimental error and the theoretical uncertainties
in the hadronic contribution is still large \cite{had}, this deviation can
be a harbinger of new physics beyond the SM. There
have already been extensive discussions on the implication 
of $\de a_\mu$ in a variety of models. These include 
SUSY theories \cite{susy}, extended Higgs sectors \cite{higgs},
extra dimensions \cite{extra}, extra gauge bosons and
leptoquarks \cite{zprime}, model independent analysis \cite{indep}, etc...

The minimal supersymmetric standard model (MSSM) is the leading
candidate for new physics beyond the SM. In the MSSM,
the  approximate one loop contribution to $\de a_\mu$ is given by \cite{susy}:
\bea
\de a_\mu^{\rm SUSY} = 13\times 10^{-10} \Bigm ( 
\frac{100 \rm GeV}{\wt m_{\rm SUSY}}\Bigm )^2 
\tan\beta\; \rm sign(\mu) \label{one}
\eea
As can be seen from the above equation, the  BNL experimental data
can easily be accommodated with positive $\mu$,
large $\tan\beta$ and light charginos, neutralinos and smuons. 

On the other hand, as one of the popular
SUSY models, the effective SUSY models \cite{eff} have been widely studied. 
In effective SUSY models, the 1st and 2nd generation sfermions are
very heavy  and only the 3rd generation sfermions are light enough to
be relevant for low  energy and collider phenomenology, thereby
easily evading the SUSY FCNC and SUSY CP problems.  
However, the light 2nd generation sleptons favored by 
$(g-2)_\mu$ \cite{susy} can put a severe 
challenge to effective SUSY models, since the 1st and 2nd generation
of sleptons are expected to be at the same scale in such models. 

In this paper we will study the implications of a large $\delta a_\mu$ in
the framework of effective SUSY models. We will show that  the
neutral CP-even Higgs, $h^0(H^0)$, together with relatively light 3rd
generation sfermions can easily explain the $\de a_\mu$ for
large $\tan\beta$ by contributing to muon MDM 
through two-loop Barr-Zee type diagrams
shown in Fig.~\ref{fig:barr_zee}. 
In our analysis, it is found that $(g-2)_\mu$ prefers 
negative soft trilinear coupling $A_{f}$ and positive $\mu$. 
It is more sensitive to the chirality flipping 
$h^0(H^0)\wt{f}_R^*\wt{f}_L$ originating both from 
F-term and soft trilinear terms
rather than chirality conserving F and D terms.

\section{Barr-Zee type diagram contribution to $(g-2)_\mu$}
\begin{figure}[t!]
\smallskip\smallskip  
\centerline{{\hskip0.352cm\epsfxsize2.9 in \epsffile{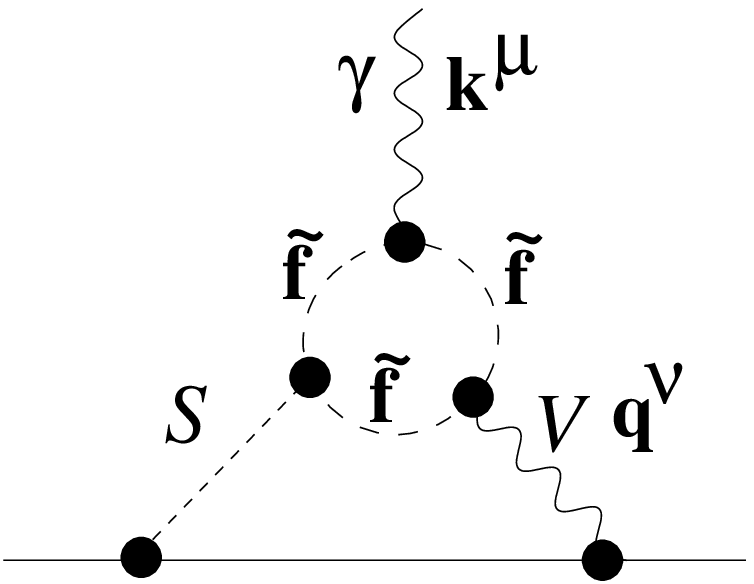}}
\hskip0.4cm}
\smallskip\smallskip\smallskip\smallskip
\caption{
Generic Barr-Zee type two-loop diagram.
}
\label{fig:barr_zee}
\end{figure}

The Barr-Zee type diagrams were first studied in the literature
\cite{barr} a long time ago.  
In particular they were shown to give large contributions to the
electric dipole moment of the electron and neutron \cite{darwin}. 
The generic Barr-Zee type diagram we are considering is
shown in Fig.~\ref{fig:barr_zee}, where $S$ denotes a generic 
scalar and $V$ a gauge boson. In the upper loop $\gamma$-$S$-$V$
 both fermions and sfermions can be exchanged. 
The fermion contribution has been studied recently
in the Two Higgs Doublet Model \cite{kong}. In this 
study, it is shown
that a sizable contribution  to 
$(g-2)_\mu$ can be obtained  for relatively light 
pseudo-scalar $A^0$ and/or 
light CP-even scalar $h^0$. This fermion contribution turns out to be 
small in the MSSM since both $A^0$ and $h^0$ are heavier 
than $\approx 90 $ GeV.   

We consider the sfermion contribution to the effective vertex
$\gamma$-$S$-$V$ in the upper loop of Fig.~\ref{fig:barr_zee}.  
We limit ourselves to the internal photon exchange ($V=\gamma$)
in the Barr-Zee diagram, since the $V=Z$ and $V=W^\pm$ diagrams
  are expected to be suppressed by their masses and also 
a small coupling in case of $V=Z$.  
We assume that CP is conserved,
and then only $(A^0 \wt f_i^*\wt f_j)_{i\neq j}$ is allowed and
 in this case only the
CP-even Higgs $S=h^0,H^0$ can contribute to $\gamma$-$S$-$V$
while $\gamma$-$A^0$-$V$ vanishes.

The Yukawa
interaction of Higgs fields with fermions or scalar fermions
can be written generically by
\begin{eqnarray}
 {\cal L}_{int} &=& -{g \lambda_f \over 2 m_W} S \overline{f} f
   -{g \lambda_{\wt{f}} \over 2 m_W} S \wt{f}^* \wt{f}.
\label{Hff}
\end{eqnarray}
With this generic coupling, and owing to electromagnetic 
gauge invariance, the effective vertex $h^0(H^0)$-$\gamma$-$\gamma$
can be written as:
\bea \Gamma^{\mu\nu}(q)
= -i  (g^{\mu\nu} q\cdot k -q^\mu k^\nu) { g e^2 \over (4 \pi)^2  m_W}
\sum_{f} N_c^{\wt{f}} \lambda_{\wt{f}} Q_{\wt{f}}^2  \int_0^1 dx
\frac{x(1-x)}{x(1-x) q^2 -m_{\wt{f}}^2}. \label{ggs}
\eea 
Here $N_c^{\wt{f}}$ is the color factor and $ Q_{\wt{f}}$ and 
$m_{\wt{f}}$ are respectively  the 
electric charge  and the mass of the internal sfermion.

When deriving the above equation, (\ref{ggs}), we kept only the
linear term in $k$ because the MDM is obtained in the soft photon limit.
Note that the Lorentz structure is manifestly gauge invariant and 
it is the only possible form factor which contributes to MDM
in the CP conserving theory.
The above vertex is connected to the external muon line by integrating
over the $h^0(H^0), \gamma$ and $\mu$ propagators. In that integration
we neglected the terms proportional to $m_\mu^2$ and kept only terms
proportional to $q^2$.
After performing the second integration, $(g-2)_\mu$ 
takes the following form: 
\bea
  \de a_\mu &=& -{\al \over 2 \pi} 
    \left( G_F m_\mu^2 \over 4 \sqrt{2} \pi^2 \right) \la_\mu^S
   \sum_{\wt f} N_c^{\wt{f}}	Q_{\wt{f}}^2 {\la_{\wt{f}} \over m_S^2}
   {\cal F}(z_{\wt{f}S}),
\label{eq:g-2}
\eea
where $z_{\wt{f}S} = m_{\wt{f}}^2/m_S^2 $, 
$\la_\mu^{\{h^0,H^0\}}={\{-\sin\al,\cos\al\}/\cos\be}$.
The loop function $ {\cal F}(z) $ is given by: 
\bea
  {\cal F}(z) &=& \int_0^1 dx \frac{x(1-x) \log{z \over x(1-x)}}{z-x(1-x)}.
\eea
The asymptotic form of ${\cal F}(z)$, which may be useful for later 
discussion, is given by:
\bea
{\cal F}(z) = \left\lbrace
\begin{array}{ll}
 -(\log z+2) \qquad & z \ll 1\\
0.344 & z= 1\\
\frac{1}{6 z}(\log z +\frac{5}{3}) & z \gg 1 .
\end{array}
\right. \label{asymp}
\eea

In the flavor eigenstates, the interactions between 
$h^0(H^0)$-$\wt{f}^*$-$\wt{f}$ are given by
\bea
   {\cal L} &=& 
      -{g \over 2 m_W \sin\be} m_t (\mu^* \sin\al+A_t \cos\al)
                  h^0 \wt{t}_R^* \wt{t}_L + h.c \nl
  && -{g \over 2 m_W \sin\be} m_t (-\mu^* \cos\al+A_t \sin\al)
                  H^0 \wt{t}_R^* \wt{t}_L + h.c \nl
  && +{g \over 2 m_W \cos\be} m_b (\mu^* \cos\al+A_b \sin\al)
                  h^0 \wt{b}_R^* \wt{b}_L + h.c \nl
  && +{g \over 2 m_W \cos\be} m_b (\mu^* \sin\al-A_b \cos\al)
                  H^0 \wt{b}_R^* \wt{b}_L + h.c,
\label{Hsfsf}
\eea
where we did not include the chirality conserving F and D-term contributions, 
which do not give any enhancement to $\de a_\mu$.
The corresponding formula for $\wt{\tau}$ interactions with $h^0(H^0)$
can be obtained by replacing $b$ by $\tau$ in (\ref{Hsfsf}).
After diagonalization of the sfermion mass matrix 
the flavor eigenstates ${\wt f}_{L,R}$
in (\ref{Hsfsf}) can be expressed in terms of the mass 
eigenstates ${\wt f}_{1,2}$. The
$\la_{\wt{f}}$ in (\ref{eq:g-2}) can be easily read from (\ref{Hsfsf}).

We stress that the quantity inside the parenthesis in (\ref{eq:g-2}) 
\bea
 { G_F m_\mu^2 \over 4 \sqrt{2} \pi^2} = 23.32 \times 10^{-10}
\eea
is the magnitude of the SM electroweak contribution to $(g-2)_\mu$.
It is also interesting to note that this is also the 
order of magnitude of $\de a_\mu$.

The two-loop expression (\ref{eq:g-2}) is suppressed by the
electroweak coupling $\al/4 \pi \sim 1/1722$. To compensate this
suppression and achieve the required magnitude to explain the experimental
deviation, some enhancement factors are required. We note that
this enhancement factors are provided by: i) the large 
$\tan\be$ enhancement in the down type Yukawa coupling 
$\la_\mu^S $ and also $\la_{\wt{f}}$, ii) large positive $\mu$ and/or 
large negative $A_f$, iii) if the splitting between the internal scalar mass 
$m_{h^0(H^0)}$ and internal sfermions mass $m_{\wt{f}}$ is large
one can have  additional enhancement from the loop 
function ${\cal F}$, which follows from (\ref{asymp}).

\section{Numerical Results}
In this section we will discuss our numerical results 
in the framework of the effective SUSY model. As stated 
in the introduction, in such models,
the 1st and 2nd generation sfermions are very heavy
while the 3rd generation sfermions can be light
without any conflict with low energy constraints. 
It follows that the  one-loop contribution to $(g-2)_\mu$
in effective SUSY model is suppressed by the heaviness of smuons.
Therefore we do not include the
one-loop contributions in our numerical analysis.
When diagonalising the mass matrix of the 3rd generation sfermions,
we neglect the small D-terms and assume that left-handed and right-handed 
soft mass parameters are degenerate, that is:
\bea 
m^2_{\wt{Q}_3} = m^2_{\wt{t}} = m^2_{\wt{b}} =  m^2_{\wt{L}_3} =
m^2_{\wt{\tau}}  \equiv \wt{m}^2.  
\eea 
Those assumptions allow  maximal mixings between left and right
chirality states.
In the absence of chirality conserving terms 
$h^0(H^0)\wt{f}_{L,R}^* \wt{f}_{L,R}$, 
the Lagrangian given in (\ref{Hsfsf}) can be
written easily in terms of the sfermion mass eigenstates by replacing
 $h^0(H^0)\wt{f}_{L,R}^* \wt{f}_{R,L}$ by 
$\pm \sin 2\theta_{\wt f}   h^0(H^0)\wt{f}_{1,2}^* \wt{f}_{1,2}$.
One can see that maximal sfermion mixings lead to 
maximal $h^0(H^0)\wt{f}_{1,2}^* \wt{f}_{1,2}$ couplings.
We also assume the universality of soft SUSY breaking trilinear couplings: 
$A_t=A_b=A_{\tau}\equiv A_0$. 
We parameterize the sfermion sector using the following 
input parameters:$\tan\beta$,  $\mu$, $A_0$ and the light 
sfermion mass $m_{{\wt f}_1}$. The heavy sfermion mass
$m_{{\wt f}_2}$ can be determined from
the above inputs.

The MSSM Higgs sector is parametrized by the mass of 
the CP-odd $M_A$ and $\tan\beta$ while the top quark mass and 
the associated squark masses enter through radiative 
corrections \cite{okada}. In our study we 
will include the leading corrections only, where the light Higgs mass
is given by:
\begin{eqnarray}
& & m^2_{h^0,H^0} =  \frac{1}{2}\Big[ m_{AZ}^2
\mp \sqrt{ m_{AZ}^4-4m_A^2 m_Z^2\, {\cos}^2\left( 2\beta \right)
-4\epsilon \left( m_A^2 \,s_\beta^2 +m_Z^2 \, c_\beta^2 \right) } \Big] \nl
& &\tan \left( 2 \alpha\right) =  \frac{m_A^2 + m_Z^2}{m_A^2 - m_Z^2 + 
\epsilon/ \cos \left( 2\beta \right)}
\, \tan \left( 2 \beta\right)
\quad\qquad  \,-\frac{\pi}{2} \le \alpha \le 0\, \; .
\end{eqnarray}
with 
\begin{eqnarray}
& & m_{AZ}^2 = m_A^2+m_Z^2+\epsilon \qquad , 
\qquad  
\epsilon = \frac{3 G_F}{\sqrt{2} \pi^2 } \frac{m_t^4}{ s_\beta^2}
\mbox{log} \left[\frac{m_{\tilde{t}_1}m_{\tilde{t}_2}}{m_t^2} \right] .
\end{eqnarray}
Our inputs are fixed in the following ranges: $2\leq \tan\beta\leq 60$, 
$100\leq  M_A \leq 250$ GeV,  
$500\leq \mu\leq 6000$ GeV,  
$-6000\leq A_0 \leq -500$ GeV,  
$100\leq  m_{\wt{f}_1} \leq 200$ GeV. 
In figures 2, 3, 4 and 5, the number beside each 
line represents the number of standard deviation from the 
central value of $\de a_\mu$.
Since we need large $LR$ mixing, we need some degree of cancellation
between $LL$ and $LR$ components to achieve the above range for
light sfermions. Therefore, all the four entries of the sfermion
mass matrices have the same order of magnitude.
In Fig.2 we show the allowed regions in the $(\tan\beta,\mu)$ and 
$(\tan\beta,A_0)$ plane. We observe that the central value 
of $\de a_\mu$ can be easily accommodated with relatively large
$\tan\beta$ and large $\mu$ or $A_0$. Even $\mu$ and $A_0$ 
around $500$ GeV, are compatible with $\de a_\mu$.  

In Fig. 3 and 4 it can be seen that light 
sfermion $m_{{\wt f}_1}$ and light 
CP-odd $M_A$ are favored both for moderate and large $\tan\beta$. 
We have checked that the dominant contribution
comes from the heavy CP-even scalar rather than light one. 
Although the former is suppressed by its mass
the enhancement due to the loop function is more important.  

In Fig. 5 we show the $\de a_\mu$ dependence on $\mu$ and $A_0$
for light CP-odd scalar and light sfermion, $M_A=m_{\wt{f}_1}=100$ GeV
for $\tan\be=50$ (left panel) and $\tan\be=35$ (right panel). For
large $\tan\be$, even $\mu \sim -A_0 \sim 0.5$ TeV can accommodate
$\de a_\mu$. For moderate $\tan\be$, we need 
at least $\mu \sim -A_0 \sim 1$ TeV
to explain $\de a_\mu$.



\section{Discussions and Conclusions}

In this paper we considered the contribution of  two-loop Barr-Zee
type diagram with scalar fermion and neutral Higgs exchange to
$(g-2)_\mu$ in the effective SUSY model.  
It turns out that only the neutral higgs ($S=h^0,H^0$)
and photon exchange ($V=\gamma$) give sizable contributions.
It is possible to produce a large enough value to
account for the discrepancy of $\de a_\mu$ between the SM prediction and
the BNL measurement.
We showed that moderate to large $\tan\be$ (30 to 50)
  and large mixing between the
left-right sfermions can easily explain $\de a_\mu$ without
invoking the lightness of the 2nd generation smuon.

We found that the sfermions should be light enough ($< 200$ GeV)
to be produced in the
future experiments, such as the Tevatron Run II or LHC.

In our scenario, since the light stop is relatively light, 
it can contribute to the FCNC process $b \to s \gamma$ through the
one-loop chargino and stop contribution. In addition, the charged Higgs
mass is of the same order
as the pseudoscalar higgs mass $M_A$, its contribution
is also relevant. 
However, to study the 
MSSM contribution to $b \to s \gamma$, we need to specify the
$SU(2)$ gaugino mass, $M_2$, which enters the chargino mass matrix.
Since the parameter space we considered is indepedent of $M_2$,
we do not further investigate $b \to s \gamma$ in our scenario.
We just note that the positive $\mu$ which is favoured by $\de a_\mu$
in our analysis
is also favoured by $B(b \to s \gamma)$.
This is because for $\mu >0$ the
chargino contribution interferes destructively with the SM and charged Higgs
contribution, which can help evade strong $B(b \to s \gamma)$
constraint.

{\bf Note Added:}
While we were finishing this work, we received a paper \cite{geng} dealing
with similar two-loop Barr-Zee type diagrams. 
However, in our paper we consider the neutral higgs and photon exchange,
whereas \cite{geng} analyses the charged Higgs and $W^\pm$ 
gauge boson contributions.
Their result seems to prefer the same sign of $\mu$ and $A_0$ as ours,
which means that their result and our result can be added constructively.
This can further increase the allowed parameter regions where BNL data
can be accommodated.

\acknowledgements
We are grateful to A. Akeroyd and P. Ko for useful discussions. 
This work was supported in part by National Science
Council  under the grants NSC 89-2112-M-002-063 and MOE CosPA Project.


\begin{figure}[t!]
\smallskip\smallskip  
\centerline{{\hskip0.352cm\epsfxsize2.9 in \epsffile{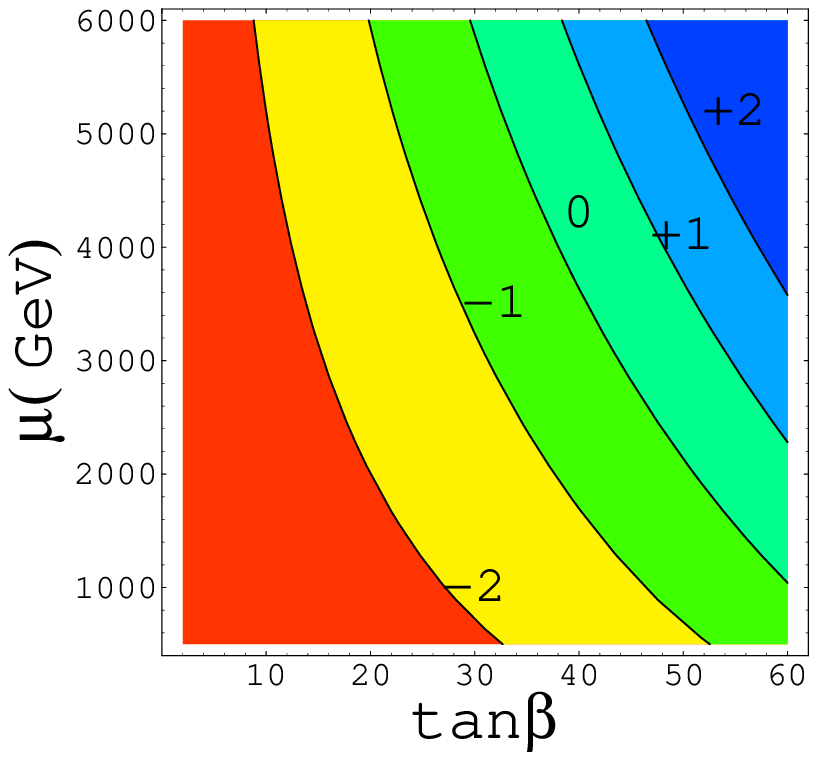}}
\hskip0.4cm
{\epsfxsize2.9 in \epsffile{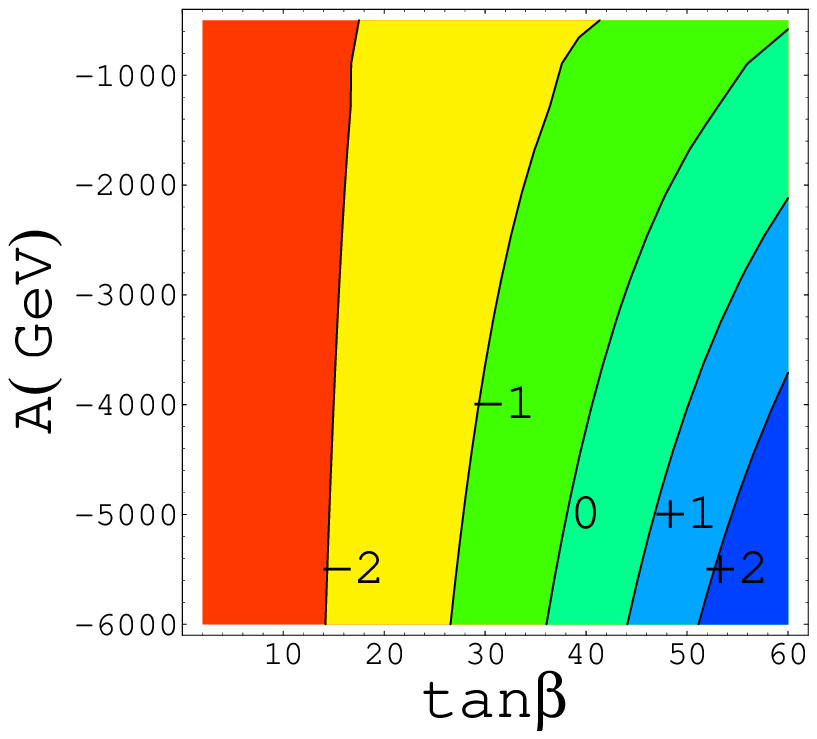}}}
\smallskip\smallskip\smallskip\smallskip
\caption{Contours of the two-loop sfermion contribution to $\de a_\mu$ in
$(\tan\be,\mu)$ plane with $M_A=100$, $m_{\wt{f}_1}=100$, $A_0=-3000$ GeV 
(left panel),and in $(\tan\be,A_0)$ plane with $M_A=100$, 
$m_{\wt{f}_1}=100$, $\mu=3000$ GeV (right panel).
}
 \label{fig:bsg2}
\end{figure}

\begin{figure}[t!]
\smallskip\smallskip  
\centerline{{\hskip0.352cm\epsfxsize2.9 in \epsffile{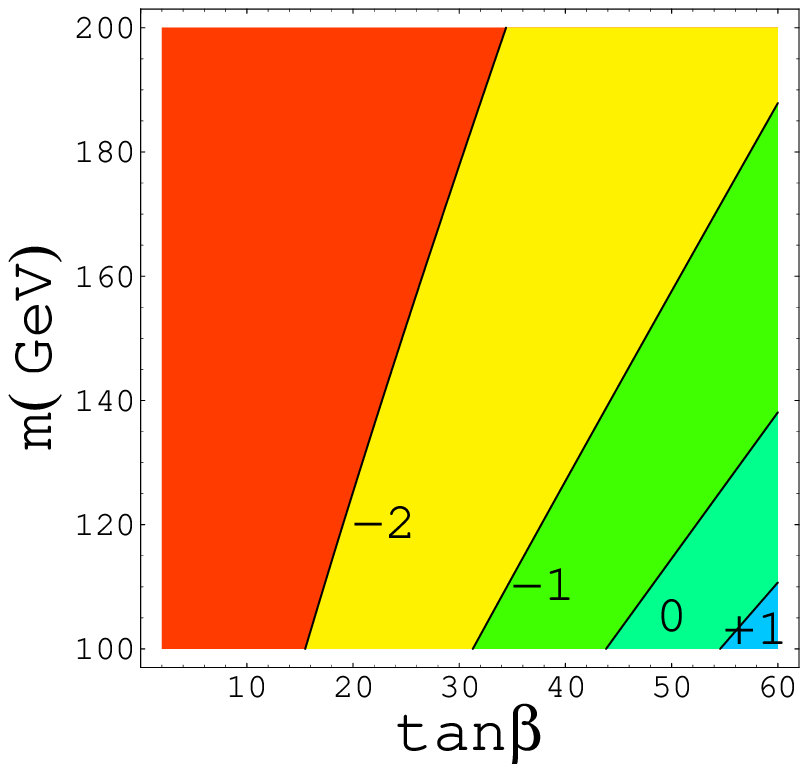}}
\hskip0.4cm
{\epsfxsize2.9 in \epsffile{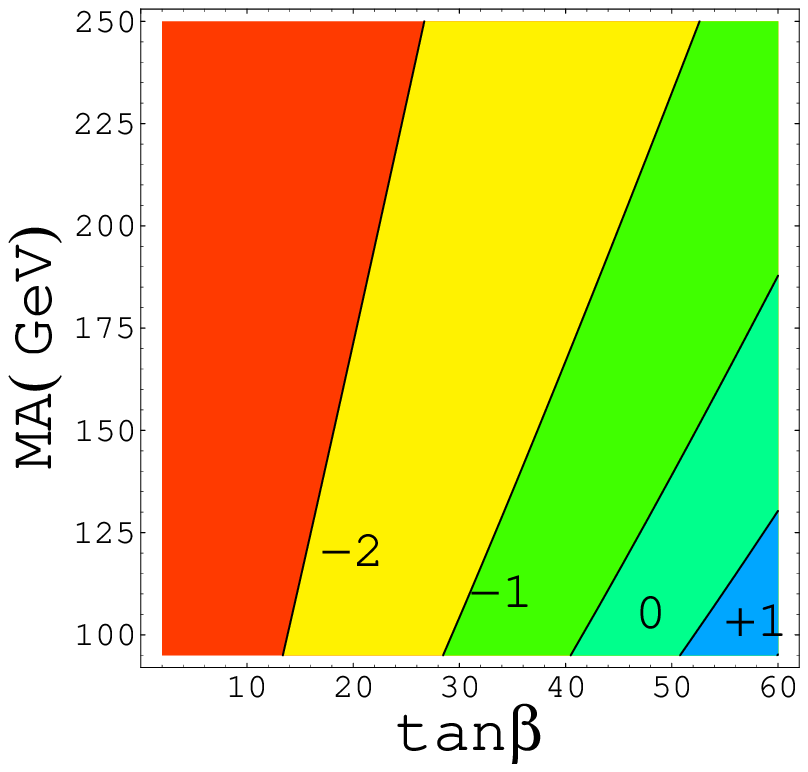}}}
\smallskip\smallskip\smallskip\smallskip
\caption{Contours of the two-loop sfermion contribution to $\de a_\mu$ in
$(\tan\be,m_{{\wt f}_1})$ plane with $M_A=100$, $\mu=3000$, $A_0=-3000$ GeV 
(left panel),and in $(\tan\be,M_A)$ plane with $A_0=-3000$, 
$m_{{\wt{f}}_1}=100$, $\mu=3000$ GeV (right panel). }
 \label{fig:bsg3}
\end{figure}

\begin{figure}[t!]
\smallskip\smallskip  
\centerline{{\hskip0.352cm\epsfxsize2.9 in \epsffile{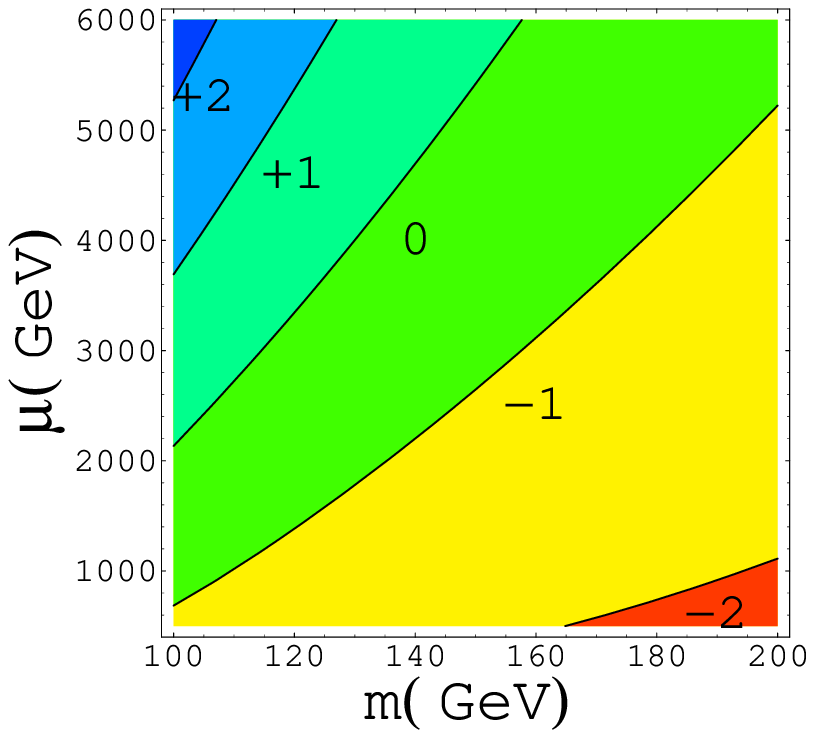}}
\hskip0.4cm
{\epsfxsize2.9 in \epsffile{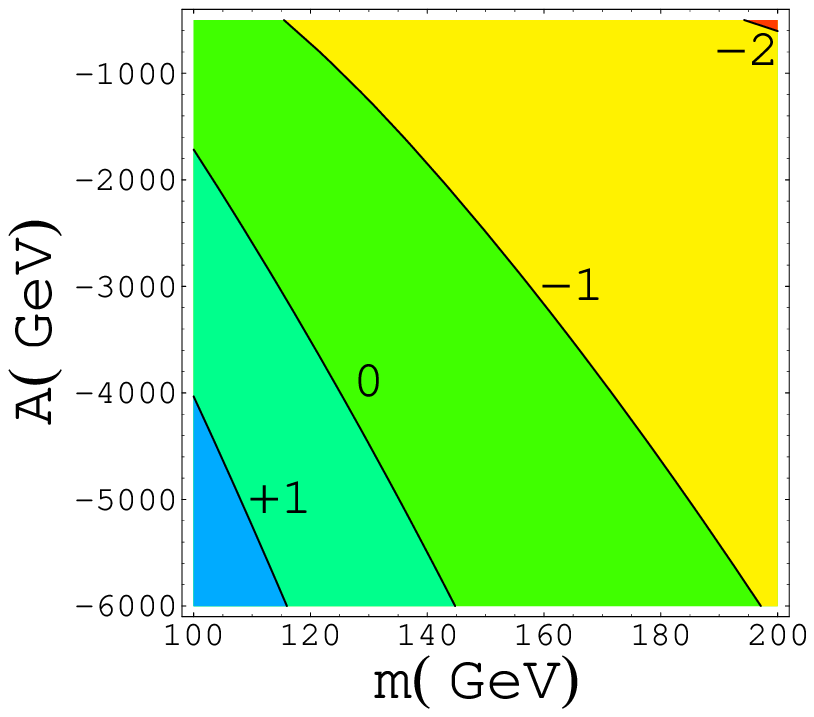}}}
\smallskip\smallskip\smallskip\smallskip
\caption{ Contours of the two-loop sfermion contribution to $\de a_\mu$ in
$(m_{\widetilde{f}_1}, \mu)$ plane with $M_A=100$, $A_0=-3000$ GeV and 
$\tan\beta=50$ 
(left panel),and in $(m_{{\wt f}_1},A_0)$ plane with $\mu=3000$, 
 $M_A=100$ GeV and $\tan\beta=50$ (right panel).
 }
 \label{fig:bsg4}
\end{figure}

\begin{figure}[t!]
\smallskip\smallskip  
\centerline{{\hskip0.352cm\epsfxsize2.9 in \epsffile{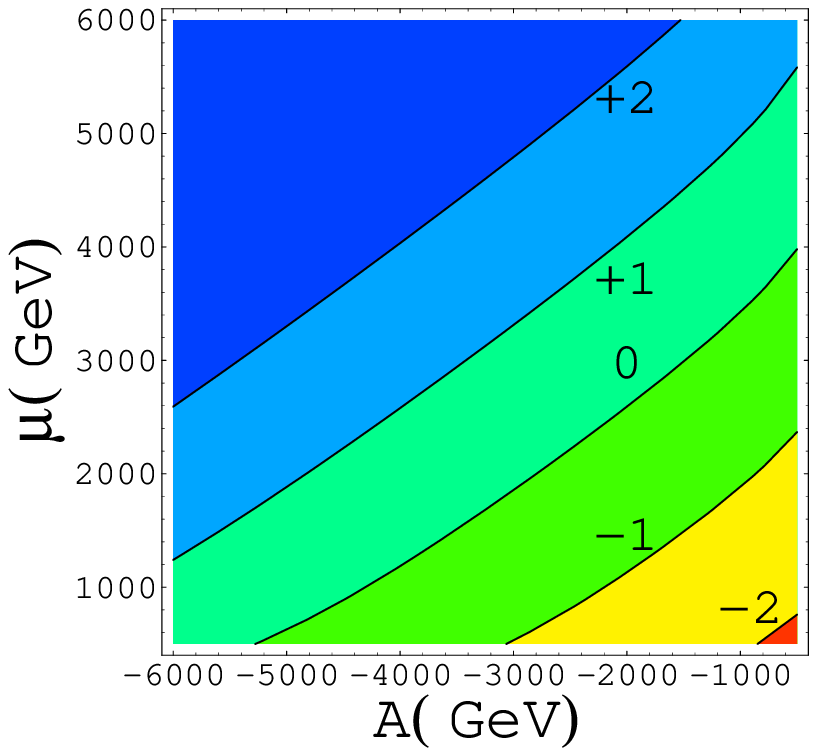}}
\hskip0.4cm
{\epsfxsize2.9 in \epsffile{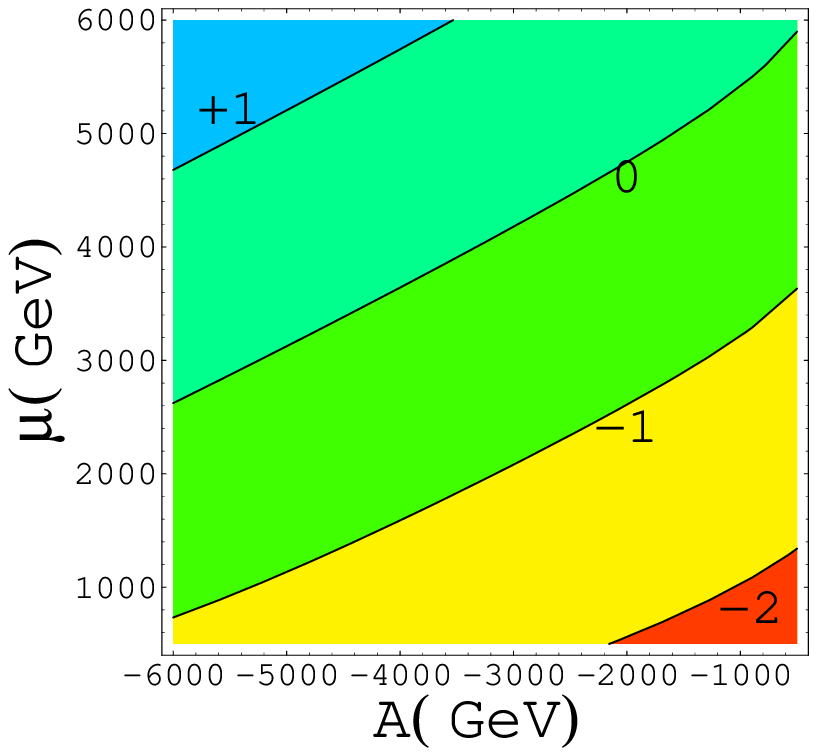}}}
\smallskip\smallskip\smallskip\smallskip
\caption{ Contours of the two-loop sfermion contribution to $\de a_\mu$ in
$(A_0, \mu)$ plane with $M_A=100$, $m_{{\wt f}_1}=100$ GeV and 
$\tan\beta=50$ 
(left panel),and in $(A_0,\mu)$ plane with $m_{{\wt f}_1}=100$, 
 $M_A=100$ GeV and $\tan\beta=35$ (right panel).}
 \label{fig:bsg5}
\end{figure}

\end{document}